# Demonstration of a Chip-based Nonlinear Optical Isolator

Shiyue Hua[1], Jianming Wen[2], Xiaoshun Jiang[1,3*], Qian Hua[1], Liang Jiang[2], and Min Xiao[1,3,4*]

[1]National Laboratory of Solid State Microstructures, College of Engineering and Applied Sciences, and School of Physics, Nanjing University, Nanjing 210093, China.

[2]Department of Applied Physics, Yale University, New Haven, Connecticut 06511, USA.

[3]Synergetic Innovation Center in Quantum Information and Quantum Physics, University of Science and Technology of China, Hefei, Anhui 230026, China.

[4]Department of Physics, University of Arkansas, Fayetteville, Arkansas 72701, USA.

*Correspondence to: jxs@nju.edu.cn; mxiao@uark.edu.

Abstract: Despite fundamentally challenging in integrated (nano)photonics, achieving chip-based light nonreciprocity becomes increasingly urgent in signal processing and optical communications. Because of material incompatibilities in conventional approaches based on Faraday effects, alternative solutions have resorted to nonlinear processes to obtain one-way transmission. However, revealed dynamic reciprocity in a recent theoretical analysis[1] has pinned down the functionalities of these nonlinear isolators. To overcome this dynamic reciprocity, we here report the first demonstration of a nonlinear optical isolator on a silicon chip enforced by phase-matched parametric amplification. Using a high-Q microtoroid resonator, we realize highly nonreciprocal transport at the 1,550 nm wavelength when waves are simultaneously launched in both forward and backward directions. Our design, compatible with current CMOS technique, yields convincing isolation performance with sufficiently low insertion loss for a wide range of input power levels. Moreover, our work evidences the possibility of designing chip-based real nonlinear isolators for information processing and laser protection[2].

Nonreciprocal photonic devices[2] with the breach of time-reversal symmetry provide crucial functionalities such as isolation and circulation in laser protection, optical signal processing, and instrumentation applications. Yet, reciprocity, as constrained by the Lorentz theorem[3], is fundamental to light transport in linear, time-invariant optical systems and holds even in rather complex ones. Although broadly used in optical communications and sensing, nonreciprocal devices are still challenging in silicon integrated photonics due to prime limitations in materials integration and device design. To break the reciprocity, a traditional means is to guide light through materials accompanying with strong magneto-optical Faraday effects[4,5]. Regardless of its versatility, this approach usually encounters severe obstruction from the miniaturization of bulky volumes and materials compatibility with mature integrated-silicon photonic platforms. Despite small footprint[6,7] could be obtained with the advanced bonding and deposition technologies, the application of an external magnetic field may deleteriously interfere with nearby optics and influence their functionalities.

The quest for alternative and more compact isolation schemes has recently garnered an immense impetus, and spawned a variety of methods by adopting different physical principles to avoid the need for the integration of magneto-optical elements. In one alternative direction, a notable effort has been made upon reproducing the effect of magneto-optics using non-magnetic structures undergoing spatiotemporal modulations[8-12] (an idea akin to the one a while ago used for nonreciprocal mode conversion in optical fibers[13]). In spite of the conceptual elegance, unfortunately, most of the reported systems[9-12] to date have to rely on complex structures, which demand operating thresholds often with fairly low nonreciprocal outputs. In contrast to linear isolators, considerable enthusiasm has been devoted to break the Lorentz reciprocity with use of various nonlinearities. Among these, to name a few, nonreciprocal light transmission has been illustrated with second-order nonlinearity[14,15], Kerr or Kerr-type nonlinearities[16,17], gain/absorption saturation[18-20], Raman amplification[21], stimulated Brillouin scattering[22], thermo-optic effect[23], and opto-acoustic effect[24] in recent progress.

Of these nonlinear schemes[14-24], asymmetric transmission contrast is typically demonstrated when a wave is injected in either forward or backward direction but never both. The lack of an experiment on the simultaneous presence of waves from both directions causes people to suspect whether these nonlinear isolators could be capable of providing complete isolation under practical operating conditions. In a very recent theoretical work, this hypothesis has been partially disproved by Shi and his coworkers[1]. Specifically, they found that for Kerr or Kerr-like nonlinearities, due to the existence of a dynamic reciprocity, nonlinear isolators of this type fail to grant any isolation for arbitrary backward-propagating noise coexisting with forward signal. Moreover, their results point out an important limitation on the use of nonlinear optical isolators for signal processing and laser protection. The discovery on dynamic reciprocity further prompts these authors to query whether such a property is generally accompanied with a nonlinear optical isolator. It is therefore fundamentally intriguing to know whether nonlinear means could be chosen to construct a *real* nonlinear optical isolator for practical applications, especially for laser protection.

To give an affirmative answer, in this Letter we present the first experimental realization of a chip-based nonlinear optical isolator in a high-Q silica microtoroid resonator[25] by exploiting phase-matched optical parametric amplification. In comparison with previous works[14-24] utilizing nonlinearities, our experiment shows explicitly remarkable isolating performance in suppressing the transmission of backward-propagating noise with launching signal fields from both directions at the same time. Imposed by the phase matching within the microcavity, the forward input signal experiences parametric amplification, while the backward input remains almost intact owing to the lack of suitable phase matching. Because of the inversion symmetry of silica, the dominant parametric process is four-wave mixing through third-order nonlinearity. In the process, two pump photons (with angular frequency $\omega_p$ and wave vector $\vec{k}_p$) are converted to one signal photon ($\omega_s$, $\vec{k}_s$) plus one idler ($\omega_i$, $\vec{k}_i$) by satisfying the phase-matching condition: $2\omega_p = \omega_s + \omega_i$ (energy conservation) and $2\vec{k}_p = \vec{k}_s + \vec{k}_i$ (momentum conservation). It is the latter momentum conservation that further removes the subtle dynamic reciprocity and enables desirable isolation for backward-traversing noise. Alternatively, this observation plays an essential role in our proof-of-principle demonstrations. The bandwidth of the parametric gain for the signal field is largely determined by the dispersion as well as the circulating optical power inside the microcavity, and it is relatively narrower than the cavity resonance linewidth in our experiment[26,27]. To keep the system stable, optical nonreciprocity is demonstrated with the dropped pump power below the threshold of the optical parametric oscillation. We note that optical parametric oscillation and its enabled Kerr frequency comb in high-Q microcavities have been studied in previous studies[28,29]. It is worth mentioning that another major challenge of our experiment is to fabricate the sample with almost no backscattering effect[30] for both pump and signal waves.

As schematically illustrated in Fig. 1a, our design consists of a high-Q silica whispering-gallery-mode microtoroid[25], fabricated on a silicon chip, and evanescently coupled to two tapered optical fibers (labelled as fiber 1 and fiber 2). The experimental setup (see Methods for details) is depicted in Fig. 1c, where the forward signal beam was seeded from port 1; while the backward signal was launched from either port 3 (in the two-fiber-coupling case) or port 2 (in the single-fiber-coupling case). The pump laser was always input through port 1. In the experiment, we first chose the two-fiber-coupled structure to interrogate optical isolation induced by phase-matched parametric amplification, as this structure permits versatile controllability and easily demonstrates nonreciprocal property. The pump and signal modes (shown in Fig. 1b) are properly selected for the microtoroid cavity with Q factors of $8.03 \times 10^7$ and $7.40 \times 10^7$ at the wavelengths of 1562.94 nm and 1556.10 nm, respectively. The wavelength difference between the pump and signal fields is about 6.84 nm, coinciding with the free spectral range (FSR) of the microcavity. The idler is also generated one FSR away from the pump mode, which makes it easier to be separated and filtered out from the pump and signal after port 2.

Before examining nonlinear optical isolation, we begin with the verification of reciprocal transmission of the forward and backward signal light by turning the pump light off. As expected, reciprocal transport (Fig. 2a) is recorded whenever the microtoroid is subject to the forward and backward signal input simultaneously or separately. The single-peak transmission spectrum observed in both directions (Fig. 2a) also confirms the absence of the backscattering inside the cavity. We then switch on the pump laser and thermally lock it into the microcavity to produce a parametric gain for the seeded signal light. Thanks to the phase-matched parametric amplification, the forward signal now returns more output from port 3 than the backward signal as long as the gain compensates the loss. Typical nonreciprocal transmission spectra are represented in Fig. 2b, where the sharp peak appearing near the spectral center of the forward original input symbolizes the maximal location of the parametric gain. Similar to other nonreciprocal light transmission enabled by resonant structures, the performance of our isolator can be well controlled by tuning a series of system's parameters, such as optical coupling rates ($\kappa_1, \kappa_2$) and dropped pump power ($P_p$). Figure 2c shows the measurement of nonreciprocal signal transmission as a function of $P_p$. Its trend indicates that the isolation ratio (blue circles) grows from 0 dB to ~18 dB by gradually increasing $P_p$. The relatively large error fluctuations are mainly due to the large uncertainty from the backward transmitted-signal measurement. This is further confirmed with the stable output in the forward configuration (red circles). During the process, the incident forward and backward signal beams were maintained with equal power of 5.6 μW, and the coupling rate $\kappa_1$ ($\kappa_2$) between fiber 1 (2) and the toroid was set at $2\pi \times 1.95$ MHz ($2\pi \times 0.18$ MHz). It is worthwhile to emphasize that this is the first measurement on a chip-based optical isolator with the coexistence of the signal light launched from both directions.

More importantly, the acquired isolation here evidently implies that the current scheme is not subject to the dynamic reciprocity.

Figure 3 is a plot of the isolation behavior versus the reflectivity. Here, the reflectivity is referred to as the fraction of the backward input signal power ($P_b$) over the total input signal power in both directions, $\eta = P_b/(P_b + P_f)$. In the experiment, the pump wavelength was thermally locked to the cavity mode with a fixed detuning. The input power was about 451.20 $\mu$W, while the dropped pump power was kept at 78 $\mu$W. The input signal power in the forward direction was set at $P_f = 0.97$ $\mu$W, while $P_b$ increases gradually from 95.57 nW to 9.57 $\mu$W. Due to the contamination in the experiment, the optical Q-factors corresponding to the pump and signal modes reduce to $4.20 \times 10^7$ and $5.79 \times 10^7$. Remarkably, reliable isolation with a ratio well above 10 dB is approachable in the range of $\eta$ between 0 and 1. The inset is a snapshot of the typical transmission spectra obtained in the forward and backward directions. To further evaluate the device performance, we have investigated the nonreciprocity in terms of the input signal powers under the condition of $P_b = P_f$ (see Fig. S3 in Supplementary Information). In this case, we fix all other parameters but only alter input powers of both forward and backward signal light. Apparent isolation can be well held with a ratio above 15 dB for $P_b = P_f$ in the range of 0.1 $\mu$W to 10 $\mu$W. In addition, the typical insertion loss as low as ~4 dB is achieved throughout the measurement (e.g. Fig. 2).

Unlike previous demonstrations[18,20,23] (which rely highly upon a microresonator asymmetrically coupled to two waveguides), optical nonreciprocity induced by the phase-matched parametric amplification can be even implementable with only a single-fiber coupling. This is understoodable if being aware of the directionality of phase matching specified by the pump-photon momentum, as it resembles an external magnetic field applied in the common magneto-optical effect. The experimental proof is implemented via removing fiber 2 (Figs. 1a and c). Using another sample with optical Q-factors of $2.07 \times 10^7$ and $5.69 \times 10^7$ at the pump and signal modes, we first test reciprocal transmission of signal fields by switching off the pump laser. The experimental data is shown in Fig. 4a, where forward and backward input signals share equal power of 1.46 $\mu$W. By turning the pump on, the nonreciprocity becomes more appreciable as the parametric gain is large enough in the cavity. The isolation trend versus the dropped pump power is displayed in Fig. 4c. As one can see, when the dropped pump power is greater than 350 $\mu$W, asymmetric transmission starts to become distinct. The typical output spectra measured at ports 2 and 1 are presented in Fig. 4b. Again, the peak arising in the forward transmission spectrum comes from the parametric amplification.

In summary, in this work we have conceptually demonstrated the possibility of designing and implementing a nonlinear optical isolator by utilizing phase-matched parametric amplification in a chip-based high-Q microtoroid resonator. The simple scheme, operated under practical conditions, explicitly proves itself to be useful for protecting a laser from harmful reflections, which is one of the most important applications of nonreciprocal devices. Another significant aspect of this design is its ability to have low insertion loss available with the gain amplification. Because the device exploits cavity resonance enhancement and phase-matched parametric amplification, the main drawback of the current work is its working range limited to narrow bandwidth in nature. Nevertheless, it is expected that the device could hold potential applications in quantum information science with use of narrowband single photons or continuous variables. Of importance, our results convey in a confident way that it is feasible to build a real nonlinear optical isolator for laser protection and optical information processing in integrated photonics. In addition, we anticipate that our work could stimulate more efforts on identifying and developing practical and magnetic-free nonreciprocal devices based on nonlinear means.

## Methods

**Optical nonreciprocity measurements.** Figure 1c presents a schematic diagram of the experimental setup for performing isolation measurements on the proposed nonlinear optical isolator. The setup looks similar to our previous works[18,20], except that it is further allowable for simultaneous injections of signal fields in both forward and backward directions. By adjusting two optical switches S1 and S2, the system is interchangeable between two coupling cases: single-fiber-coupling and two-fiber-coupling.

**Two-fiber-coupling case.** For the experiment with the two-fiber-coupled structure, S1 and S2 were switched into ②. The forward signal was launched into the microcavity through port 1 and was dropped out at port 3. After passing an optical fiber circulator (C2), two optical couplers (Coupler5 and Coupler6), and a tunable bandpass filter (TBF1), the dropped forward signal emitted from port 3 is measured by a photodetector (D1). The forward signal exiting from port 2 was routed to S2, Coupler7, TBF3 and D4 for measurement. Correspondingly, the backward signal was injected from port 3 and dropped out of the cavity from port 1 followed by C1, Filter2 and D3 for analysis. D6 was used to measure the coupling rate between fiber 2 and the microcavity. During the measurement, the pump field was always launched from port 1 and its transmission at port 2 was detected by D5 after traversing S2 and Coupler7.

**Single-fiber-coupling case.** By switching S1 and S2 into ① as schematically shown in Fig. 1c, the system is in the single-fiber-coupling case. That is, the microtoroid resonator is only coupled with fiber 1. Fiber 2 was moved away via a nano-positioner. The forward signal beam was launched into the toroid through port 1 and its output from port 2 was directed to S2, S1, C2, Coupler5, Coupler6, TBF1, and D1 for detection. The backward signal light was incident through port 2 and its output from port 1 was directed to C1, TBF2, and D3 for measurement. The transmitted pump light from port 2 was measured by D2 after passing S2, S1, Coupler5, and Coupler6.

For the experiment on investigating nonreciprocal transmission by launching the signal either in the forward configuration or the backward as done in previous research[14-24], on the other hand, one only needs to simply switch off one path of the signal propagation through a variable optical attenuator (VOA 4 or VOA3).

**Acknowledgements**

This research was supported by the National Basic Research Program of China (2012CB921804), the National Natural Science Foundation of China (Nos. 61435007, 11574144 and 11321063), the Natural Science Foundation of Jiangsu Province, China (BK20150015), the Fundamental Research Funds for the Central Universities. J.W. and L.J. acknowledge the funding supports from the ARO MURI, ARL, AFOSR MURI, Alfred P. Sloan Foundation, and David and Lucile Packard Foundation.

**Author contributions**

X.J. and M.X. conceived the idea and supervised the experiment. S.H. implemented the experiment with the help from Q.H. and analyzed the data with contributions from J.W. and L.J. All authors participated in the discussion of the project. X.J., J.W. and S.H wrote the manuscript with contributions from all authors.

**Additional information**

Supplementary information is available in the online version of the paper. Reprints and permissions information is available online at www.nature.com/reprints. Correspondence and requests for materials should be addressed to X.J. or M.X.

**Competing financial interests**

The authors declare no competing financial interests.

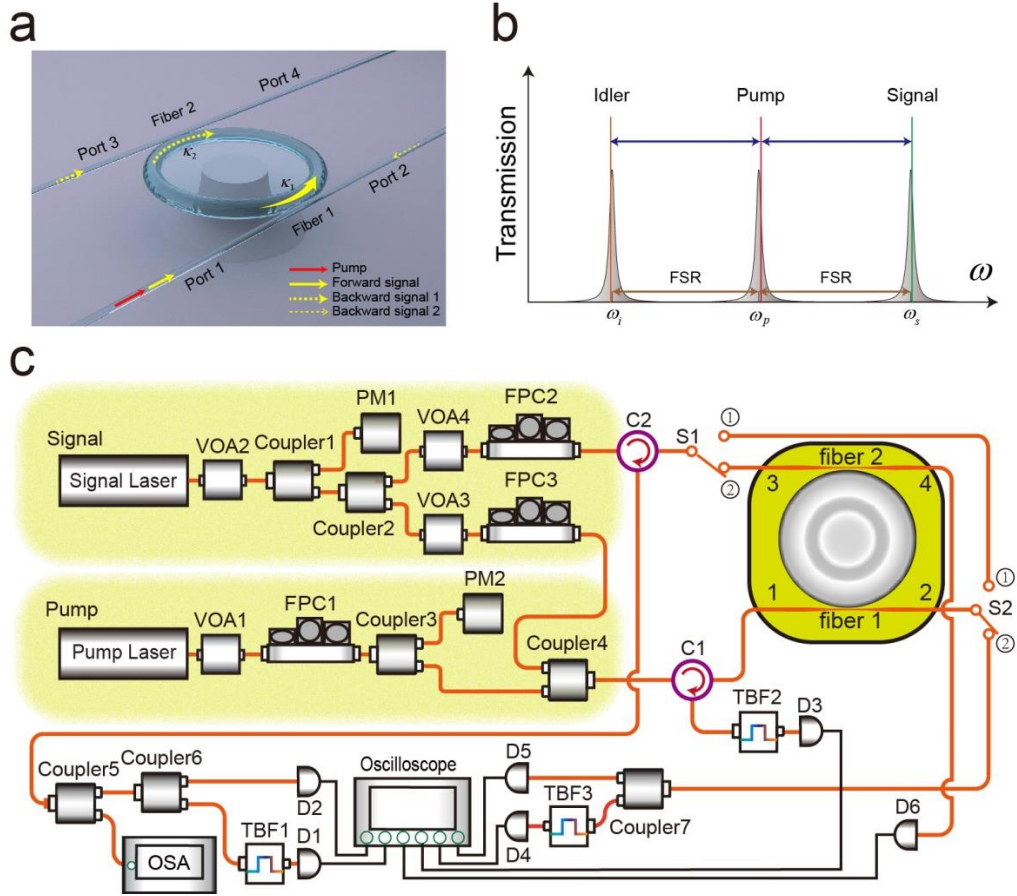

**Figure 1 | Nonlinear optical isolator based on a high-Q whispering-gallery-mode silica microtoroid resonator on a chip. a**, 3D schematic of forward (solid yellow arrow) and backward (dashed yellow arrow) propagation configurations based upon signal inputs at ports 1 and 3 in the two-fiber-coupling case (at ports 1 and 2 in the single-fiber-coupling case). **b**, Frequency spectral representation of the pump, signal, and idler waves involved in four-wave mixing, whose occurrence only appears in the forward direction due to phase matching. The red line represents the cw pump laser. The green and yellow lines denote, respectively, the parametric-amplified forward signal and generated idler waves. **c**, Schematic of the experimental setup. C: optical fiber circulator; D: photodetector; FPC: fiber polarization controller; OSA: optical spectrum analyzer; PM: power meter; S: optical switch; TBF: tunable bandpass filter; VOA: variable optical attenuator.

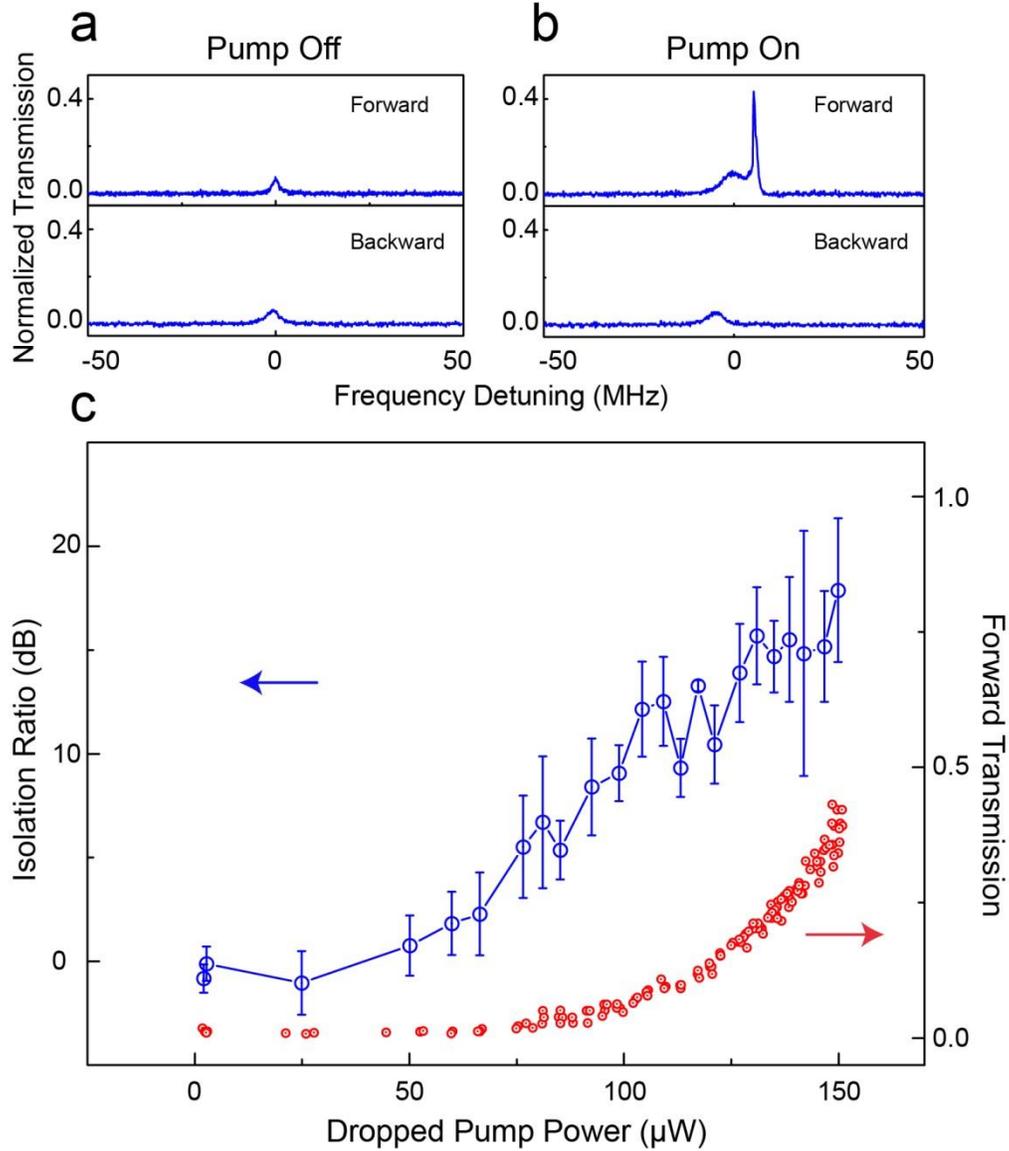

**Figure 2 | In simultaneous presence of equal forward and backward signal power of 5.6 $\mu$W, optical isolation performance of the device versus dropped pump power in the two-fiber-coupling case. a**, With pump off, as expected, reciprocal transmission is obtained. The single peak appearing in both forward and backward transmission spectra indicates negligible backscattering effect in the microcavity. **b**, With pump on, typical asymmetric transmission spectra are observed when the parametric gain compensates the loss. The sharp peak in forward transmission marks the center of the parametric gain. **c**, Measured optical isolation ratio as a function of the dropped pump power (blue circles). The normalized forward signal transmission (red circles) with a smooth behavior suggests that the isolation uncertainty be mostly owing to the backward transmission. Other parameters: $\kappa_1 = 2\pi \times 1.95$ MHz, and $\kappa_2 = 2\pi \times 0.18$ MHz.

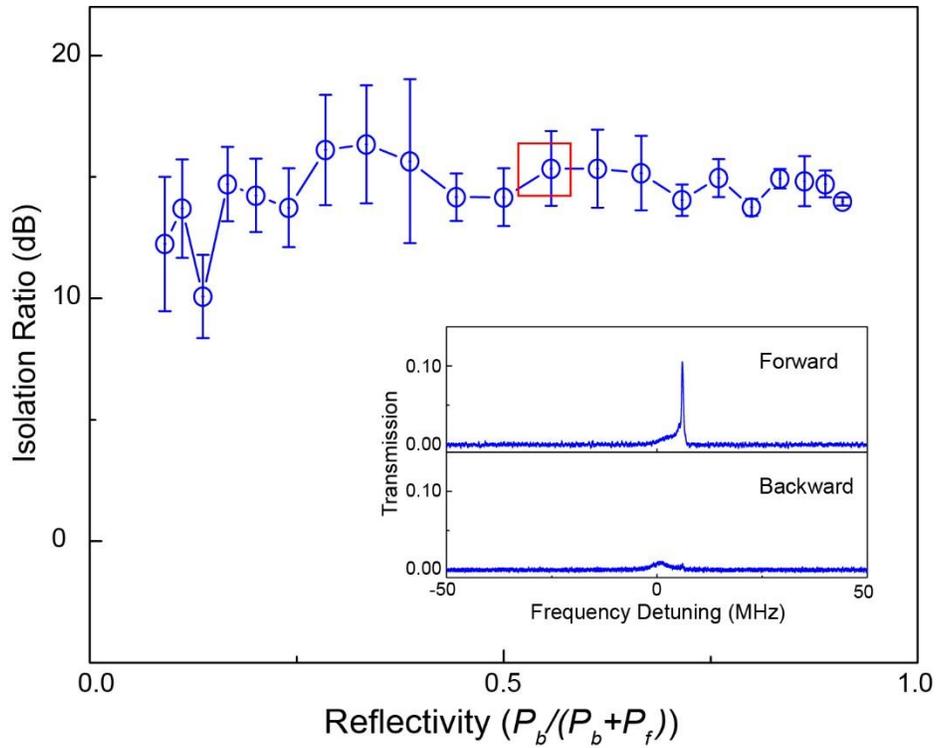

**Figure 3 | Optical isolation in terms of reflectivity with fixed forward input signal power of 0.97 $\mu$W in the two-fiber-coupling case.** The backward input signal power is increased from 95.57 nw to 9.57 $\mu$W. The inset shows the transmission spectra of the marked point with a backward signal power of 1.20 $\mu$W. Other parameters: $\kappa_1 = 2\pi \times 0.29$ MHz, and $\kappa_2 = 2\pi \times 0.05$ MHz.

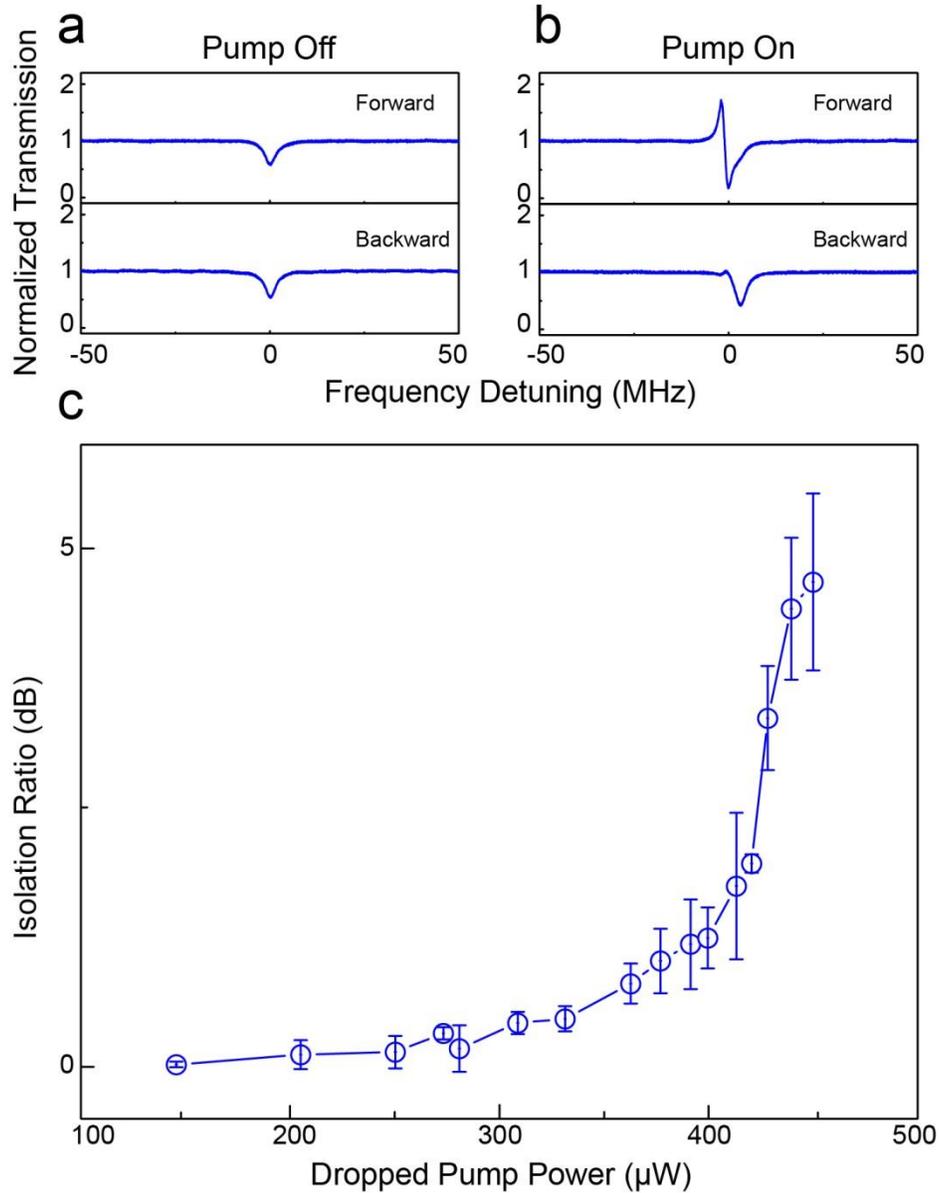

**Figure 4 | In simultaneous presence of equal forward and backward signal power of 1.458 μW, optical nonreciprocal performance of the device versus dropped pump power in the single-fiber-coupling case. a**, Reciprocal transmission is expected with pump off. **b**, With pump on, typical nonreciprocal transmission spectra are recorded as the forward signal experiences phase-matched parametric amplification but the backward does not. **c**, The optical isolation is characterized only as a function of the dropped pump power. Other parameters: $\kappa_1 = 2\pi \times 0.4$ MHz.